\documentclass[aps,prc,twocolumn,showpacs,groupedaddress]{revtex4}

\usepackage{graphicx}
\usepackage{dcolumn}
\usepackage{bm}
\def\lsim{\lower.5ex\hbox{$\; \buildrel < \over \sim \;$}}
\def\gsim{\lower.5ex\hbox{$\; \buildrel > \over \sim \;$}}

\begin{document}


\title{Cross sections relevant to gamma-ray line emission in solar flares:
$^3$He-induced reactions on $^{16}$O nuclei}


\author{V. Tatischeff, J. Duprat, J. Kiener, M. Assun\c{c}\~ao, A. Coc, 
C. Engrand, M. Gounelle, A. Lefebvre, M.-G. Porquet, N. de S\'er\'eville and 
J.-P. Thibaud}
\affiliation{Centre de Spectrom\'etrie Nucl\'eaire et de Spectrom\'etrie de
Masse, IN2P3-CNRS and Universit\'e Paris-Sud, F-91405 Orsay Cedex, France}

\author{C. Bourgeois, M. Chabot, F. Hammache and J.-A. Scarpaci}
\affiliation{Institut de Physique Nucl\'eaire, IN2P3-CNRS and Universit\'e
Paris-Sud, F-91406 Orsay Cedex, France}


\date{\today}

\begin{abstract}
Gamma-ray production cross sections have been measured for gamma-ray
lines copiously emitted in the $^3$He bombardment of $^{16}$O nuclei:
the 937, 1042 and 1081~keV lines of $^{18}$F and the 1887~keV line of
$^{18}$Ne. Four Ge detectors with BGO shielding for Compton suppression were
used to measure the angular distributions of the gamma-rays. The excitation
functions have been obtained for $^3$He bombarding energies from 3.7 to 36
MeV. Total cross sections are tabulated for calculations relevant to 
gamma-ray astronomy. The importance of these lines as diagnosis for 
the presence and properties of accelerated $^3$He in solar flares is 
discussed in light of the measured cross sections. 
\end{abstract}


\maketitle

\section{Introduction}

Solar flares have been historically divided into two classes, impulsive and 
gradual, based on the time duration of their soft X-ray emission (see 
\cite{rea99}). They are, however, also well characterized by the abundances
of the energetic particles accelerated in these events and detected in the
interplanetary space. In particular, impulsive flares are generally
associated with enormous enhancements of accelerated $^3$He: the measured
$^3$He/$^4$He abundance ratios are frequently 3--4 orders of magnitude
larger than the corresponding value in the solar corona and solar wind where
$^3$He/$^4$He$\sim$5$\times$10$^{-4}$ (see \cite{mas02}). These solar
energetic particle events have been the subject of many experimental and 
theoretical investigations aiming at the identification of the resonant 
acceleration mechanism causing the $^3$He enrichments (see 
\cite{pae03} and references therein).

Independent information on the accelerated particle composition and energy
spectrum, as well as on solar ambient abundances, density and temperature,
can be obtained from observations of the gamma-ray line emission
produced by nuclear interactions of the energetic particles with the solar
atmosphere (see \cite{sha01}). The recent launch of the {\it RHESSI}
satellite offers, in particular, the possibility of high-resolution
spectroscopic analyses of this emission \cite{lin00}. Mandzhavidze {\it et
al.} \cite{man97} pointed out that the reaction
$^{16}$O($^3$He,$p$)$^{18}$F$^*$ leads to strong line emission at 937, 1042
and 1081 keV, which can be used as diagnosis for the presence and
properties of accelerated $^3$He. The 937 keV line, as well as a significant 
line feature at $\sim$1.02~MeV have already been identified in data of the 
moderate-resolution gamma-ray spectrometers {\it SMM}/GRS and {\it CGRO}/OSSE 
\cite{sha98}. Analysis of these detections suggested that the flare-averaged
$^3$He/$^4$He abundance ratio could be $\sim$0.1 and that in some flares
$^3$He/$^4$He$\sim$1 \cite{sha98,man99}. 

However, these analyses were based on gamma-ray production cross sections
evaluated from partial laboratory data (see \cite{koz02}): excitation
functions for the production of the 937, 1042 and 1081 keV lines from the 
reaction $^{16}$O($^3$He,$p$)$^{18}$F$^*$ were measured for laboratory 
$^3$He energies ranging from 2.6 to 4 MeV only \cite{heg80}; at higher 
energies, the gamma-ray cross sections were estimated from data on $^{18}$F 
total production, obtained through the detection of the $\beta^+$ decays of 
this isotope ($T_{1/2}$=109.77~m). 

We have measured the cross sections for the production of the 937, 1042 and
1081 keV lines in $^3$He+$^{16}$O reactions, for 40 $^3$He energies ranging
from 3.7 to 36 MeV. In addition, we have considered the production of the
1887 keV line from deexcitation of the first excited state of $^{18}$Ne,
which is populated by the reaction $^{16}$O($^3$He,$n$)$^{18}$Ne. Indeed, we
suggest that the detection in solar flares of this relatively strong line 
could allow one to obtain valuable information on the energy spectrum of the 
accelerated $^3$He, by comparing its intensity to that of the 937 keV 
line. This is discussed in section IV, together with the overall relevance of
these measurements for gamma-ray spectroscopy of solar flares. The experiment 
and the data analysis are described in sections II and III, respectively. 

\section{Experimental procedure}

\begin{figure*}
\includegraphics[width=14.cm]{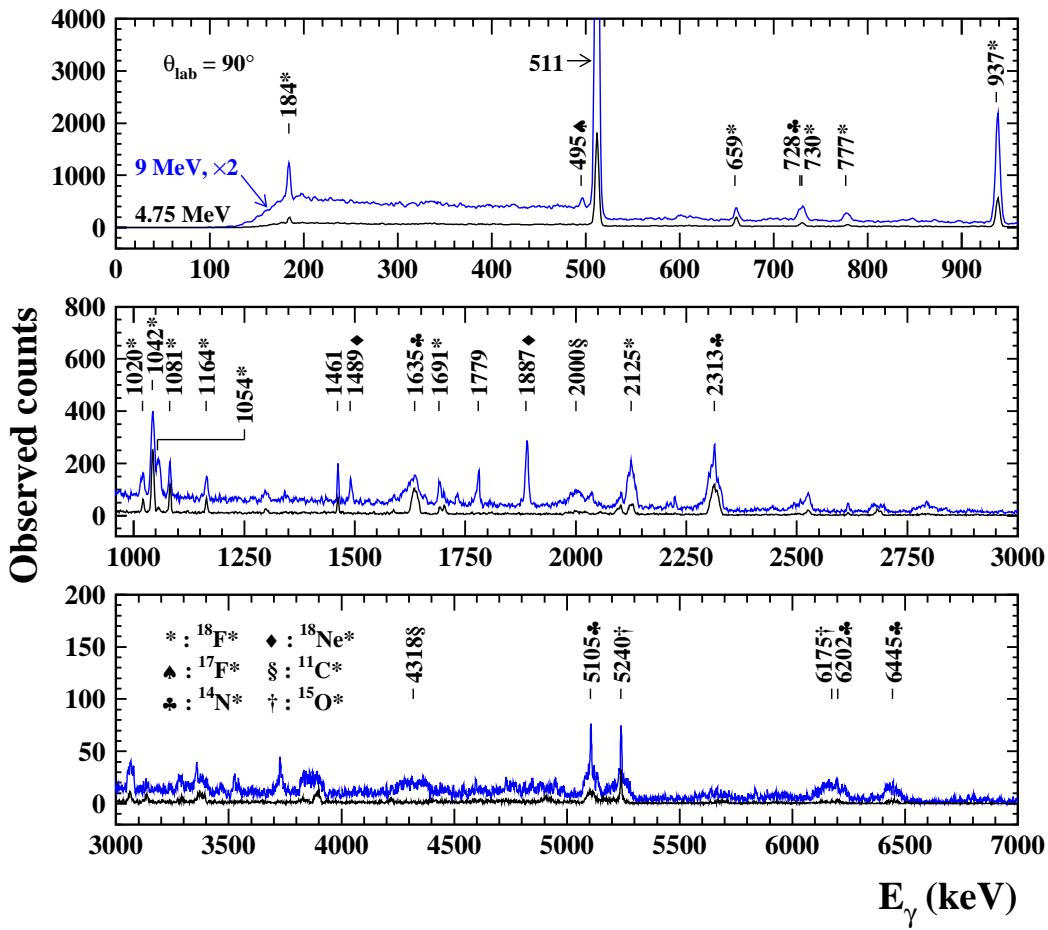}%
\caption{Observed gamma-ray spectra with the Ge detector at
$\theta_{lab}$=90$^\circ$ for $^3$He-particle bombardments at $E_{beam}$=4.75
and 9~MeV of the tantalum-oxide target of 800~$\mu$g/cm$^2$ evaporated on a
carbon foil of 90~$\mu$g/cm$^2$. Most of the lines arise from $^3$He-induced
reactions on $^{16}$O and $^{12}$C target nuclei (see text). The most intense
of these lines are labelled with their nominal energies and the excited
nuclei from which the gamma-rays are emitted. Almost all of the unlabelled 
weak lines between 2500 and 4000 keV are due to cascade transitions following 
the $^{16}$O($^3$He,$p$)$^{18}$F$^*$ and $^{12}$C($^3$He,$p$)$^{14}$N$^*$ 
reactions. The strong 511 keV line mainly arises from the $\beta^+$ decay of 
the $^{11}$C, $^{15}$O, $^{17}$F, $^{18}$F and $^{18}$Ne ground states. The 
background line at 1461 keV arises from the decay of ambient $^{40}$K 
($T_{1/2}$=1.3$\times$10$^9$~y). The width of this line, 3.5~keV full width
at half maximum, reflects the energy resolution of the detector. The line at 
1779~keV is due to deexcitation of the first excited state of $^{28}$Si and 
is believed to arise from secondary-particle interactions in the aluminum 
walls of the reaction chamber.  
\label{fig1}}
\end{figure*}

The experiment was performed at the Tandem accelerator of the IPN-Orsay. 
Together with the cross sections for gamma-ray production from 
$^3$He+$^{16}$O reactions, we studied in the same experiment the reaction 
$^{24}$Mg($^3$He,$p$)$^{26}$Al, which is relevant to the production of 
$^{26}$Al in the early solar system \cite{dup03}. 
The $^3$He-induced reactions on $^{16}$O and $^{24}$Mg were successively
studied at each beam energy by means of a multiple target holder, which was
shifted in the vertical direction with a geared engine remote-controlled from
the data acquisition room. For each $^3$He energy, the beam position and
spot size were first controlled optically with an alumina foil mounted on the
target holder. Then, the magnesium- and oxygen-containing targets were
successively placed in the beam. The gamma-ray production cross sections were
measured for 40 beam energies, in small steps of 250 to 500 keV from 3.7 to 
15 MeV and at 18, 20, 21, 25, 30 and 36 MeV. The beam current was integrated 
in a Faraday cup located 1.4~m behind the target. Typical beam intensities 
were 10-20~nA. 

\begin{figure}
\includegraphics[width=8.6cm]{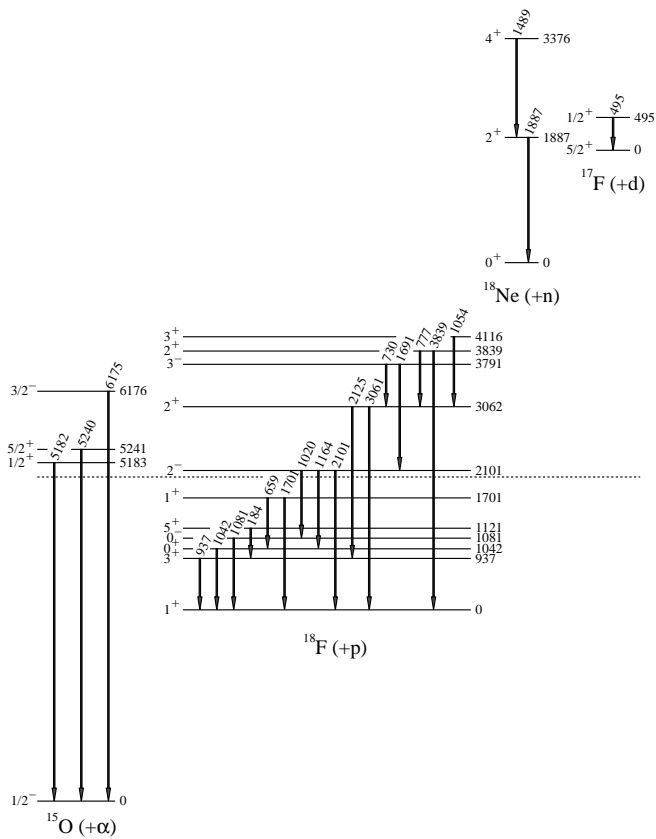}%
\caption{Partial level schemes of nuclei produced in $^3$He+$^{16}$O
reactions, giving the gamma transitions in interest. They are drawn
according to the $Q$-values of the reactions 
$^{16}$O($^3$He,$\alpha$)$^{15}$O, $^{16}$O($^3$He,$p$)$^{18}$F, 
$^{16}$O($^3$He,$n$)$^{18}$Ne and $^{16}$O($^3$He,$d$)$^{17}$F: 4914, 2032, 
-3196 and -4893~keV, respectively. The dashed line 
corresponds to the rest-mass energy of the $^3$He+$^{16}$O system.
\label{fig2}}
\end{figure}

Several target foils were used for the $^3$He+$^{16}$O measurements. From
$E_{beam}$=3.7 to 7.25 MeV and at 9 and 10 MeV, we used a tantalum-oxide
target (chemical composition Ta$_2$O$_5$) of 800~$\mu$g/cm$^2$ evaporated on
a carbon foil backing of 90~$\mu$g/cm$^2$. Above 7.25 MeV, we essentially
used two Ta$_2$O$_5$ targets of 1.5~mg/cm$^2$ evaporated on carbon foils of
130~$\mu$g/cm$^2$, as well as two self-supporting Mylar foils (chemical
composition C$_{10}$H$_8$O$_4$) of 3.213~mg/cm$^2$ (23~$\mu$m) for
$E_{beam}$=12.5, 18, 21, 25, 30 and 36 MeV. The thicknesses of the
Ta$_2$O$_5$ targets were chosen for the $^3$He energy loss in the targets
to be similar to the beam energy steps
of 250 keV below 10 MeV, in order to avoid significant gaps in the
excitation functions. The reason of using Mylar targets
instead of Ta$_2$O$_5$ targets at the highest energies was to avoid a too
intense production of neutrons and gamma-rays from reactions with the Ta
nuclei. Repeated measurements at the same energies allowed to ensure the 
stability of the various targets. After the experiment, the thickness of one 
of the Ta$_2$O$_5$ targets was checked by a Rutherford backscattering 
measurement performed at the ARAMIS accelerator of the CSNSM with a 
$^4$He beam of 3.085~MeV energy. From this measurement and the observed 
consistency of cross section data measured with different targets at the same 
energies, the uncertainty on the target thicknesses was estimated to be 
$\pm$10\%.  

The gamma-rays were detected with four large volume, high purity Ge detectors
with BGO shielding for Compton suppression \cite{nol94}. The four
detectors were placed horizontally in front of 2~mm-thick aluminum-windows of
the reaction chamber and at the laboratory angles $\theta_{lab}$=90$^\circ$,
112.5$^\circ$, 135$^\circ$ and 157.5$^\circ$ with respect to the beam
direction. The Ge crystals were located at about 37~cm far from the
target. The detection efficiencies were measured at three different times by
placing at the target position four radioactive sources calibrated in
activities to 1.5\%: $^{60}$Co, $^{88}$Y, $^{137}$Cs and $^{152}$Eu. These
sources provide a dozen of gamma-ray lines of relatively high intensities
from 122 to 1836 keV. The uncertainties in the absolute efficiencies were
determined from these three measurements to be $\pm$7\% for the detectors at
112.5$^\circ$, 135$^\circ$ and 157.5$^\circ$, and $\pm$15\% for the detector
at 90$^\circ$ due to an additional absorption of the gamma-rays in the target
holder. The efficiencies at 1333~keV ($^{60}$Co) were
(1.67$\pm$0.25)$\times$10$^{-4}$, (2.28$\pm$0.16)$\times$10$^{-4}$,
(1.90$\pm$0.13)$\times$10$^{-4}$ and (1.97$\pm$0.14)$\times$10$^{-4}$, for
the detectors at 90$^\circ$, 112.5$^\circ$, 135$^\circ$ and 157.5$^\circ$, 
respectively. 

Conventional electronic and computing techniques were used to translate the
detector output signals into pulse height spectra. The dead time of the
system was precisely determined with a set of pulsers which were fed into the 
Ge detector preamplifiers and into a scaler that was triggered by the data
acquisition system. We assessed an error of $\pm$3\% in the dead-time
correction, which was $\sim$20\% for most of the measurements, but as high as
$\sim$60\% for the highest beam energies. 

Sample gamma-ray spectra are shown in Fig.~1 for the Ta$_2$O$_5$ target of
800~$\mu$g/cm$^2$ on the carbon backing. We see a variety of lines produced by
$^3$He-induced reactions on $^{16}$O and $^{12}$C nuclei. For this target, the 
$^{16}$O-to-$^{12}$C number ratio is N($^{16}$O)/N($^{12}$C)=1.2$\pm$0.2, 
which is lower by a factor of $\sim$2 than the corresponding
abundance ratio in the solar atmosphere
[N($^{16}$O)/N($^{12}$C)]$_{\rm{sol}}$=2.4 \cite{and89}. The first and second 
escape peaks are almost invisible and the Compton background is strongly 
suppressed by the BGO anticoincidence system. For
$E_{beam}$=4.75 MeV (lower spectrum in Fig.~1), all significant lines arise
from the exothermic reactions $^{16}$O($^3$He,$p\gamma$)$^{18}$F,
$^{12}$C($^3$He,$p\gamma$)$^{14}$N and
$^{16}$O($^3$He,$\alpha\gamma$)$^{15}$O. 

Apart from the strong 511~keV emission, the most intense line is at 937 keV
and originates from the deexcitation of the first excited level in 
$^{18}$F (Fig.~2). This line is relatively narrow because the mean-life of 
the 937 keV state, $\tau$=67.6$\pm$2.5~ps \cite{til95}, is greater than the 
slowing down time of the recoiling $^{18}$F ions in the target material, such 
that the gamma-ray emission happens predominantly near rest. The reaction 
$^{16}$O($^3$He,$p\gamma$)$^{18}$F also produces four lines, at 1020, 1042, 
1054 and 1081 keV, which could contribute to the $\sim$1.02~MeV line emission 
feature detected in solar flares with {\it SMM}/GRS and {\it CGRO}/OSSE 
\cite{sha98}. 

The strongest lines from the reaction
$^{12}$C($^3$He,$p\gamma$)$^{14}$N are at 2313, 1635 and 5105 keV. It was not
possible to extract reliable cross section data for $^3$He+$^{12}$C reactions
in this experiment, because the Ge detectors were not shielded enough from
gamma-rays produced in the graphite material of the Faraday cup. In solar 
flares, these three lines are expected to be also significantly
excited by inelastic scattering of protons and $\alpha$-particles on
ambient $^{14}$N \cite{koz02}. 

The most significant line from the reaction
$^{16}$O($^3$He,$\alpha\gamma$)$^{15}$O is at 5240~keV corresponding to the
deexcitation of the second excited level in $^{15}$O (Fig.~2). The line at 
5182~keV from the first excited level in $^{15}$O was observed to be 
significantly weaker.

For $E_{beam}$=9 MeV, three additional reactions can be identified in the 
gamma-ray spectrum shown in Fig.~1: the reaction
$^{16}$O($^3$He,$d$$\gamma_{495}$)$^{17}$F, the reaction
$^{12}$C($^3$He,$\alpha\gamma$)$^{11}$C producing two broad lines at 2000
and 4318~keV and the reaction $^{16}$O($^3$He,$n\gamma$)$^{18}$Ne producing
two narrow lines at 1489 and 1887~keV (Fig.~2). We see that the latter line 
is relatively strong, which motivated us to determine its production cross
section. The relevance of this line for solar flare physics is discussed in 
section IV. 

\section{Data analysis and results}

The cross sections for the production of the 937, 1042 and 1081 keV lines of
$^{18}$F and the 1887 keV line of $^{18}$Ne
have been derived from the yields of the full-energy peaks in the energy 
calibrated spectra. We used a Monte Carlo simulation of the gamma-ray line 
production in the targets (see \cite{kie01}) to better estimate the peak 
shapes for the four detection angles. The background yields were determined 
by interpolation of the count rates below and above the peaks. For the
937 and 1887 keV lines, the systematic error due to the peak shape
determination and the background subtraction was estimated to be comparable to
the statistical error. For the weaker line at 1081 keV, this systematic
error was typically 5\% for the detectors at 90$^\circ$, 112.5$^\circ$ and
135$^\circ$, and 10-20\% for the detector at 157.5$^\circ$. Photopeak
integrations were indeed more uncertain for this latter detector, because of 
a higher gamma-ray background induced by nuclear reactions in the Faraday cup.
The case of the 1042 keV line was complicated by the proximity of the
1054 keV line (see Fig.~1). For $E_{beam}$$>$5 MeV, these two lines were not
fully resolved from each other and we estimated their relative intensities
from fits by two Gaussian shaped lines. The associated systematic error was 
5-15\% for 5$<$$E_{beam}$$<$12~MeV and 15-35\% above 12 
MeV, as the two lines were almost impossible to distinguish for the highest 
beam energies. The total error in the gamma-ray line yields were obtained 
from the linear sum of this systematic error with the usual statistical
error. 

\begin{figure}
\includegraphics[width=8.cm]{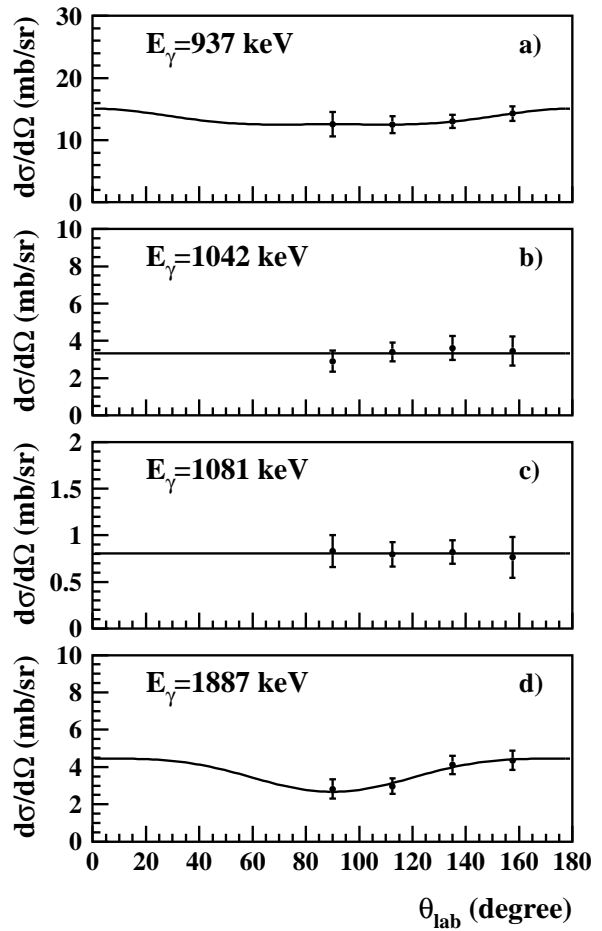}%
\caption{Examples of angular distributions of gamma-rays obtained at an
$^3$He-beam energy $E_{beam}$=9 MeV. The solid lines are 
Legendre-polynomial fits to the data (see eq.~1). The error bars
only contain the angular-dependent uncertainties and are 
obtained from a quadratic sum of the error in the number of detected 
gamma-rays and the error in the detection efficiency (see text). 
\label{fig3}}
\end{figure}

Examples of measured angular distributions of gamma-rays are shown in Fig.~3. 
The 1042 and 1081 keV line emissions are isotropics, because the gamma-rays 
depopulate 0$^+$ and 0$^-$ levels, respectively. The 937 and 1887 keV lines 
are $E2$ transitions \cite{til95} and complete angular distributions were 
obtained by fitting the measured differential cross sections by the 
Legendre-polynomial expansion:
\begin{equation}
{d\sigma \over d\Omega}(\theta_{lab})=a_0 + a_2 Q_2 P_2(\cos \theta_{lab})
+ a_4 Q_4 P_4(\cos \theta_{lab})~,
\end{equation}
where $ P_l (\cos \theta_{lab})$ is the Legendre polynomial of order $l$ and
$Q_2$ and $Q_4$ are the attenuation coefficients, which can be calculated
analytically from the geometry of the detection setup (see \cite{fer65}). We
have estimated the corrections arising from the laboratory-to-center-of-mass 
transformation and from the finite beam spot size to be negligible in this 
experiment. The total cross section is simply given by 
\begin{equation}
\sigma = 4 \pi a_0~.
\end{equation}

In Fig.~4-7, we show the excitation functions obtained from the fits of the
measured differential cross sections. Energies at which cross sections and
Legendre-polynomial coefficients are quoted are laboratory $^3$He energies at
the middle of the targets, the horizontal error bars corresponding to the
target thicknesses. Stopping powers were calculated with the computer code
TRIM \cite{ziegl}. The vertical error bars correspond to the
1$\sigma$-errors in the coefficients of the Legendre-polynomial fits, except
for the 1042~keV line for which we took into account an additional error
due to the contribution of the reaction 
$^{16}$O($^3$He,$n$)$^{18}$Ne($\beta^+$)$^{18}$F$^*$ (see below). In the 
determination of absolute cross sections, errors in the target thicknesses 
(10\%), the beam current integration (5\%) and the dead-time correction (3\%) 
must also be included. The resulting overall uncertainties are 13-14\% in 
$\sigma$(937~keV), 13-23\% in $\sigma$(1042~keV), 14-16\% in 
$\sigma$(1081~keV) and 13-17\% in $\sigma$(1887~keV). 

\begin{figure}
\includegraphics[width=8.6cm]{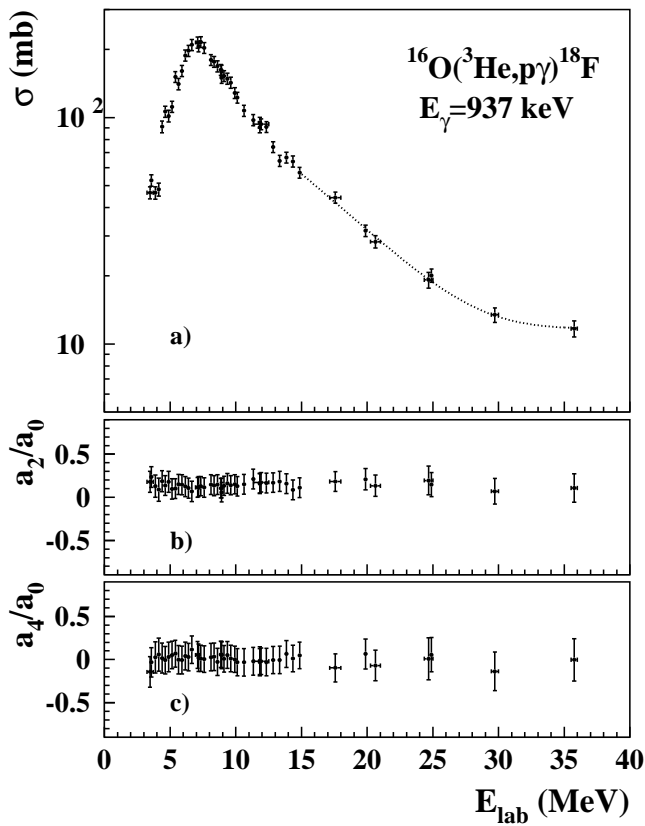}%
\caption{Total cross section (panel {\it a}) and coefficients of the 
Legendre-polynomial fits to the angular distributions (panels {\it b} and 
{\it c}) for the reaction $^{16}$O($^3$He,$p\gamma_{\rm{937}}$)$^{18}$F, as 
functions of the laboratory $^3$He energies at the middle of the targets. 
The horizontal error bars correspond to the target thicknesses. The vertical 
error bars are the errors in the Legendre-polynomial fits. The dotted line in 
panel ({\it a}) shows the polynomial fit to the data above $E_{lab}$=15 MeV 
used to calculate the average values of the cross section at high energies 
(see text). 
\label{fig4}}
\end{figure}

We see that the excitation functions vary with energy in a relatively smooth
fashion. Only the cross section for the ($^3$He,$n$) channel (Fig.~7) shows
resonances corresponding to compound nuclear effects (see below). 
The lack of fine structures in the excitation functions is partly due to the 
use of relatively thick targets (160-330~keV thickness for 
$E_{beam}$=3.7-12~MeV). 

The highest cross section is found for the reaction
$^{16}$O($^3$He,$p\gamma_{\rm{937}}$)$^{18}$F, which reaches a maximum of
216$\pm$28~mb at 7.3~MeV (Fig.~4). The angular distributions of the
937~keV gamma-rays were found to be slightly anisotropic (see also Fig.~3)
and to have little dependence on the $^3$He bombarding energy: the $a_2$/$a_0$
coefficient ratio is nearly constant and slightly positive (its mean value is
0.14$\pm$0.12) whereas all values of $a_4$/$a_0$ are compatible with zero. 
Because the 937~keV level of $^{18}$F has a relatively long mean-life, 
$\tau$=67.6$\pm$2.5~ps, and high magnetic dipole moment, 
$\mu$=+1.68$\pm$0.15 \cite{til95}, it is possible that its nuclear alignment 
was partly lost by hyperfine interactions in the target prior to the gamma 
emission. 

\begin{figure}
\includegraphics[width=8.6cm]{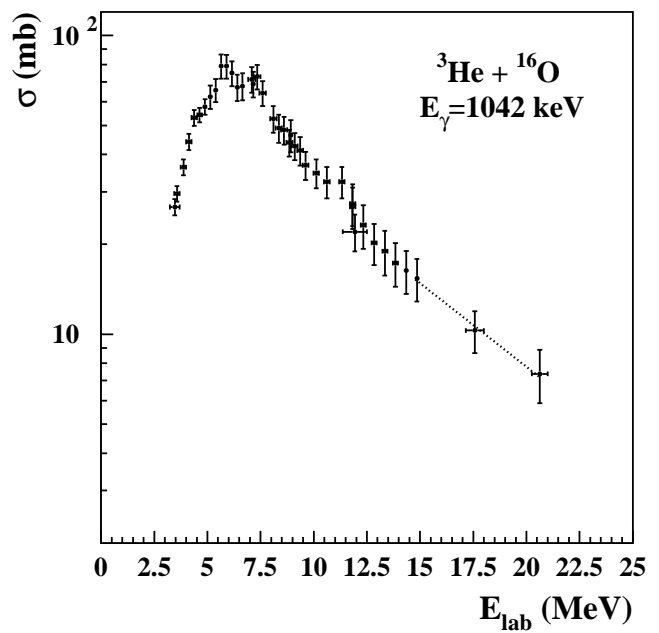}%
\caption{Total cross section for the production of 1042 keV gamma-rays from
$^3$He+$^{16}$O reactions. The 1042 keV state of $^{18}$F is mainly
populated by the reaction $^{16}$O($^3$He,$p$)$^{18}$F$^*_{\rm{1042}}$, with a
contribution $<$25\% from the reaction
$^{16}$O($^3$He,$n$)$^{18}$Ne($\beta^+$)$^{18}$F$^*_{\rm{1042}}$ (see text).
Vertical errors bars are obtained from a quadratic sum of the errors in the
Legendre-polynomial fits and the errors ($\leq$10\%) in the correction for lost
($^3$He,$n$) events. The dotted line shows the exponential fit used to 
calculate the averaged values of the cross section above 15 MeV. 
\label{fig5}}
\end{figure}

In solar flares, production of the 1042~keV line includes a contribution from
the reaction $^{16}$O($^3$He,$n$)$^{18}$Ne($\beta^+$)$^{18}$F$^*$ for $^3$He
energies above the reaction threshold $E_{\rm{^3He}}^{\rm{Thres}}$=3.80~MeV. 
For each $\beta^+$ decay of $^{18}$Ne, 0.0783$\pm$0.0021 gamma-rays of 
1042~keV are emitted \cite{til95}. The half-life of $^{18}$Ne, 
$T_{1/2}$=1672$\pm$8~ms, is too short for the ($^3$He,$n$) and ($^3$He,$p$) 
channels to be distinguished in solar flares. In this experiment, however, a 
small correction of the 1042~keV data had to be made to take into account the 
nondetection of some ($^3$He,$n$) events due to the escape from the targets 
of some recoiling $^{18}$Ne ions prior to decay. To evaluate this correction, 
we first estimated the total cross section for the reaction 
$^{16}$O($^3$He,$n$)$^{18}$Ne from the differential cross
sections $(d\sigma/d\Omega_{cm})$ measured by Adelberger and McDonald 
\cite{ade70} at $E_{lab}$=9, 9.5, 10.5, 11.5 and 12.5 MeV, and by Nero
{\it et al.} \cite{ner81} at $E_{lab}$=13.8, 16.1 and 17.8 MeV, as well as 
from the $^{16}$O($^3$He,$n\gamma_{\rm{1887}}$)$^{18}$Ne excitation function 
measured in this experiment (Fig.~7). From this estimate, we found that 
the contribution of $^{18}$Ne production to the total 1042~keV line emission 
is lower than 25\%. We then used the measured differential cross sections
$(d\sigma/d\Omega_{cm})$ \cite{ade70,ner81} to calculate the energy
distributions of the recoiling $^{18}$Ne ions and performed TRIM simulations
for each target to evaluate the fraction of escaping ions. The most
important correction was found for the Ta$_2$O$_5$ target of 
1.5~mg/cm$^2$ bombarded at $E_{beam}$=15~MeV. For this data point, we
estimated that $\sim$60\% of produced $^{18}$Ne have escaped from the target 
and the corresponding correction was applied by multiplying the measured 
cross section by a factor 1.11$\pm$0.11. Corrections for the Mylar targets 
and for the Ta$_2$O$_5$ targets bombarded at lower energies were less 
important. The excitation function for the production of the 1042~keV line
is shown in Fig.~5. 

\begin{figure}
\includegraphics[width=8.6cm]{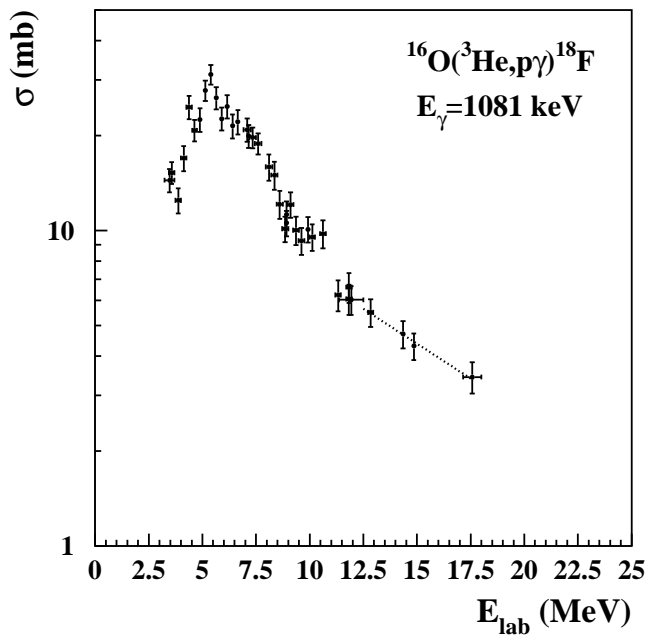}%
\caption{Total cross section for the reaction
$^{16}$O($^3$He,$p\gamma_{\rm{1081}}$)$^{18}$F. Errors bars are as explained
in Fig.~4. The dotted line shows the exponential fit used to calculate the 
average values of the cross section above 13 MeV. 
\label{fig6}}
\end{figure}

The lowest cross section measured in this experiment is that of the reaction
$^{16}$O($^3$He,$p\gamma_{\rm{1081}}$)$^{18}$F, which reaches a maximum of
only 31$\pm$4~mb at 5.4~MeV (Fig.~6). The $\beta^+$ decay of $^{18}$Ne also
populates the 1081~keV state of $^{18}$F, but with a negligible branching
ratio of (2.07$\pm$0.28)$\times$10$^{-5}$ \cite{til95}. 

\begin{figure}
\includegraphics[width=8.6cm]{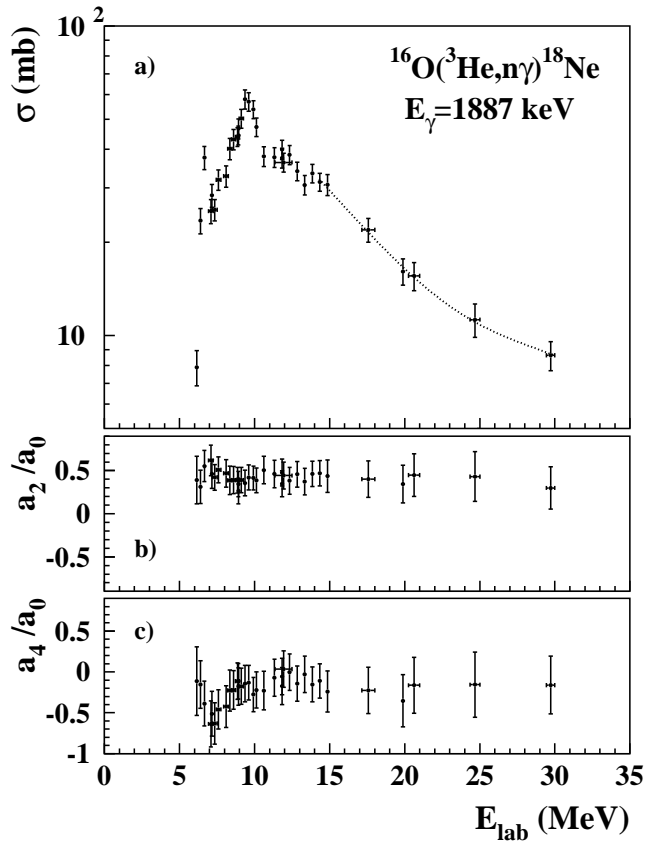}%
\caption{Same as Fig.~4 but for the
$^{16}$O($^3$He,$n\gamma_{\rm{1887}}$)$^{18}$Ne reaction. 
\label{fig7}}
\end{figure}

Results for the $^{16}$O($^3$He,$n\gamma_{\rm{1887}}$)$^{18}$Ne reaction are
shown in Fig.~7. A broad resonancelike structure dominates the excitation
function at $E_{lab}$$\approx$9.5~MeV, which corresponds to an excitation 
energy of $\approx$16.4~MeV in the compound nucleus. This broad resonance has 
already been reported in the yields of neutron production \cite{ade70}. A 
narrower resonance is also observed at $E_{lab}$$\approx$6.5~MeV, which could
correspond to the 13.8~MeV level of $^{19}$Ne \cite{til95}. The
Legendre-polynomial coefficients show a characteristic structure at
$E_{lab}$$\approx$7~MeV, with a significant deep for $a_4$/$a_0$. Above
$\sim$8.5~MeV, the $a_2$/$a_0$ and $a_4$/$a_0$ coefficient ratios do not
significantly vary with energy and their mean values are 0.40$\pm$0.17 and
-0.15$\pm$0.23, respectively. 

Heggie {\it et al.} \cite{heg80} have measured the excitation functions for
the production of the 937, 1042 and 1081 keV lines in the reaction
$^{16}$O($^3$He,$p$)$^{18}$F$^*$ up to 4 MeV.
Comparison of their results with the cross sections we obtained at the lowest
energies show an overall agreement of $\sim$20\%. The results of our
gamma-ray production measurement can also be compared with the cross section 
for $^{18}$F total production from $^3$He+$^{16}$O reactions, which has been 
extensively studied by an activation method (see \cite{fit77}). The reported 
excitation function has a very similar overall shape to the one of the sum 
of the four excitation functions measured in our experiment and is 
$\sim$20\% higher. The difference is due to the direct production of $^{18}$F 
and $^{18}$Ne in their ground states, as well as to their production in 
excited levels above 1081~keV ($^{18}$F) and 1887~keV ($^{18}$Ne) decaying 
directly to their ground states. 

In order to provide data in a convenient form for gamma-ray astronomy,
we have averaged the measured cross sections over 1-MeV energy intervals up
to 15 MeV and over 5-MeV energy intervals at higher laboratory energies. The
results are given in Table~1. Below 15 MeV, average values were generally
obtained from linear interpolations between adjacent data points. At higher
energies, we performed fits to the cross section data by polynomial or
exponential curves (see Fig.~4-7). We estimated the uncertainty induced by
this procedure to be generally negligible as compared with the overall errors
in the measured cross sections, except for $\sigma$(1887~keV)
at $E_{lab}$=6 and 7~MeV. For these two values, we quadratically added  
errors of 25\% and 10\%, respectively, due to the resonance at 
$\approx$6.5~MeV (Fig.~7). 

%
\begin{table}
\caption{Cross sections for the production of gamma-ray lines in
$^3$He+$^{16}$O reactions. The cross sections are averaged over 1-MeV-wide
bins centered at the indicated $^3$He energies up to 15 MeV and
over 5-MeV-wide bins for $E_{lab}$=18, 23, 28 and 33 MeV. 
Errors are discussed in the text. 
\label{tab1}}
\begin{ruledtabular}
\begin{tabular}{ccccc}
$E_{lab}$ & $\sigma$(937 keV) & $\sigma$(1042 keV) 
& $\sigma$(1081 keV) & $\sigma$(1887 keV) \\
(MeV) & (mb) & (mb) & (mb) & (mb) \\
\hline
 3 &  21\footnotemark[1]  &   9.4\footnotemark[1]  &   9.3\footnotemark[1]  
 &   0 \\
 4 &  60  $\pm$   8  &  41  $\pm$   5  &  17  $\pm$   3  &   0  \\
 5 & 118  $\pm$  15  &  60  $\pm$   8  &  26  $\pm$   4  &   0  \\
 6 & 172  $\pm$  22  &  75  $\pm$  11  &  24  $\pm$   3  & 7.0  $\pm$ 2.0 \\
 7 & 212  $\pm$  27  &  70  $\pm$  11  &  21  $\pm$   3  &  31  $\pm$   5 \\
 8 & 187  $\pm$  24  &  56  $\pm$   9  &  17  $\pm$   2  &  35  $\pm$   5 \\
 9 & 157  $\pm$  20  &  44  $\pm$   7  &  11  $\pm$   2  &  49  $\pm$   7 \\
10 & 127  $\pm$  16  &  35  $\pm$   6  & 9.6  $\pm$ 1.4  &  51  $\pm$   7 \\
11 & 102  $\pm$  13  &  32  $\pm$   5  & 7.9  $\pm$ 1.2  &  38  $\pm$   5 \\
12 &  94  $\pm$  12  &  24  $\pm$   5  & 6.2  $\pm$ 1.0  &  38  $\pm$   5 \\
13 &  71  $\pm$   9  &  20  $\pm$   4  & 5.4  $\pm$ 0.9  &  33  $\pm$   5 \\
14 &  66  $\pm$   9  &  17  $\pm$   3  & 4.8  $\pm$ 0.8  &  32  $\pm$   4 \\
15 &  56  $\pm$   7  &  15  $\pm$   3  & 4.4  $\pm$ 0.7  &  30  $\pm$   4 \\
18 &  41  $\pm$   5  &  10  $\pm$   2  &   -             &  21  $\pm$   3 \\
23 &  23  $\pm$   3  &   -             &   -             &  13  $\pm$   2 \\
28 &  15  $\pm$   2  &   -             &   -             & 9.4  $\pm$ 1.6 \\
33 &  12  $\pm$   2  &   -             &   -             &   - \\
\end{tabular}
\end{ruledtabular}
\footnotetext[1]{From Ref.~\cite{heg80}, fig.~2. The error is
$\sim$15\%, including an uncertainty of 12\% in the target thickness.}
\end{table}

\section{Discussion}

\begin{figure*}
\includegraphics[width=12.cm]{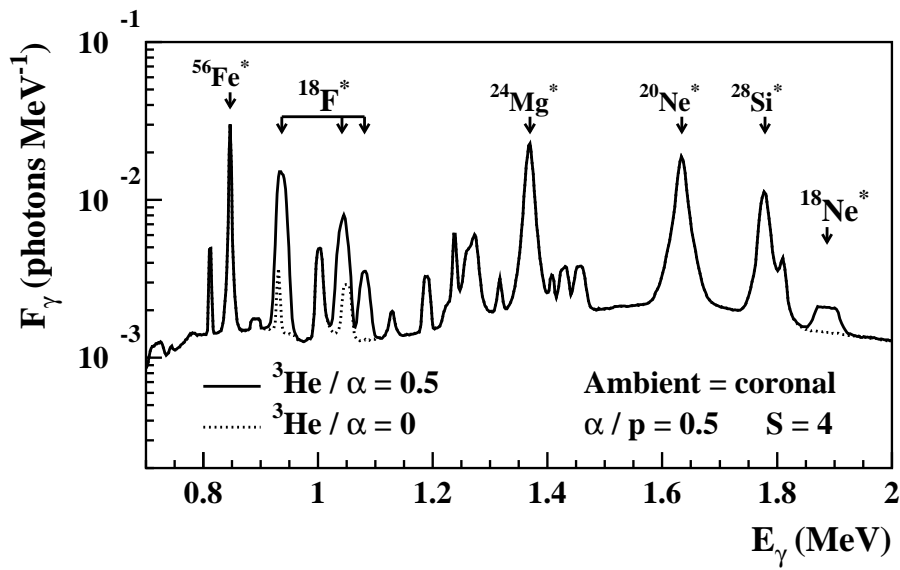}%
\caption{Calculated prompt gamma-ray line spectra in the 0.7--2 MeV region
produced by accelerated particles (mainly protons, $^3$He and
$\alpha$-particles) interacting with the solar atmosphere (see text).
{\it Solid line}: accelerated $^3$He/$\alpha$=0.5; {\it dotted line}:
$^3$He/$\alpha$=0. The fast particle source spectrum is
$Q(E) \propto E^{-S}$ with $S$=4, normalized to 1
proton of energy $E$$>$30 MeV. The 4 lines produced in $^3$He+$^{16}$O 
reactions, as well as the 4 strongest lines produced by accelerated protons 
and $\alpha$-particles are labelled with the excited nuclei from which the 
gamma-rays are emitted. 
\label{fig8}}
\end{figure*}

To discuss the expected emission of the $^3$He-produced lines in solar 
flares, we show in Fig.~8 two calculated gamma-ray line spectra, assuming  
an accelerated particle composition with or without $^3$He. We used the
code developed by Ramaty {\it et al.} \cite{ram79,koz02}, in which we 
incorporated the cross section values of Table~1. At higher $^3$He energies, 
we extrapolated the four cross sections assuming the same energy dependence 
as estimated by Kozlovsky {\it et al.} \cite{koz02} for the 
$^{16}$O($^3$He,$p$)$^{18}$F$^*$ reactions. In practice, we added two values 
for $\sigma$(937~keV), 3.4~mb at 60~MeV and 1.6~mb at 90~MeV, two values for 
$\sigma$(1042~keV), 3.6~mb at 30~MeV and 0.9~mb at 60~MeV, one value for 
$\sigma$(1081~keV), 1~mb at 30~MeV, and two values for $\sigma$(1887~keV), 
2~mb at 60~MeV and 0.9~mb at 90~MeV. In Fig.~8, these extrapolations 
account for $<$3\% of the $^3$He line strengths. 

The calculations were performed in a thick target interaction  model, in
which the accelerated particles produce the nuclear reactions while they slow 
down in the solar atmosphere. Gamma-ray line spectroscopic analyses of solar 
flare emissions have shown that the composition of the gamma-ray production 
regions is generally close to coronal (see \cite{ram95}): the abundances of 
the low FIP (first ionization potential) elements Mg, Si and Fe relative to 
the high FIP elements C and O are enhanced by a factor of 4--5 \cite{rea94} 
in comparison with photospheric abundances. We took the coronal abundances 
from Ref.~\cite{rea94}, except for He/H for which we adopted the 
helioseismological value He/H=0.084 \cite{ric98}. For the accelerated 
particles, we used an impulsive-flare average composition obtained from solar 
energetic particle measurements \cite{rea95}, except for $\alpha$/$p$ for 
which we used an abundance ratio of 0.5. Such a relatively high concentration 
of accelerated $\alpha$-particles has been proposed to explain the intense 
$\alpha$-$^4$He fusion lines observed in several flares (see 
\cite{sha98,man99}). The two gamma-ray spectra shown in Fig.~8 were calculated 
for $^3$He/$\alpha$=0 ({\it dotted line}) and $^3$He/$\alpha$=0.5 ({\it solid
line}). The latter value is typical of the $^3$He enrichment of the solar
energetic particles accelerated in impulsive flares (see \cite{rea94}). For
the source energy spectrum of the fast ions, we took for all species a power
law in kinetic energy per nucleon, $Q(E) \propto E^{-S}$, with $S$=4 (see
\cite{ram96}). 

We see in Fig.~8 that the 937 keV line is relatively strong in comparison
with other lines produced by accelerated protons and $\alpha$-particles.
However, we also see from the gamma-ray spectrum calculated without
accelerated $^3$He, that analysis of this line can be complicated by the
proximity of another line at 931 keV, which is produced in the reaction
$^{56}$Fe($p$,$pn$)$^{55}$Fe$^*$ \cite{man99,koz02}. In Fig.~9a,
we show the fluence ratio of the 931 keV line to the 937 keV line,
as a function of the power law spectral index of the accelerated particle
energy spectrum. We used for $\sigma$(931~keV) the values supplied
in table~A3 of Ref.~\cite{koz02}. The relative contribution of the 931~keV
line rapidly diminishes for increasing $S$, because this line is produced
for proton energies above the reaction threshold $E_p$=11.4~MeV, whereas
$\sigma$(937~keV) peaks at significantly lower $^3$He energy per nucleon,
$E_{\rm{^3He}}\cong$2.4~MeV/nucleon (see Table~1 and Fig.~4). Given the
assumed coronal abundance ratio $^{56}$Fe/$^{16}$O=0.16 and the adopted
accelerated particle composition with $^3$He/$p$=0.25, the 931~keV line
fluence is $<$5\% of that of the 937~keV line for $S$$>$4. 

\begin{figure}
\includegraphics[width=8.6cm]{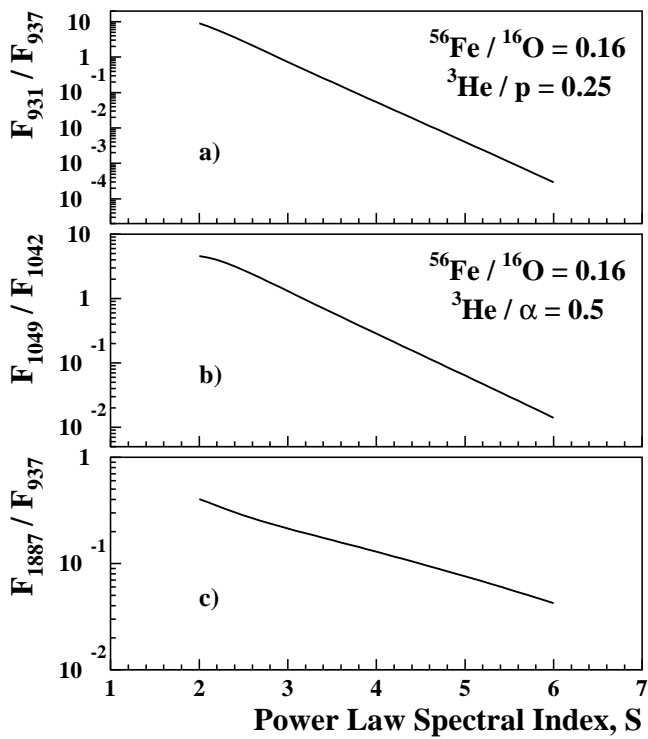}%
\caption{Calculated line fluence ratios in a thick-target interaction model
as a function of the power-law spectral index of the accelerated particles.
The lines at 931 keV ({\it a}) and 1049 keV ({\it b}) are produced by proton
and $\alpha$-particle reactions on $^{56}$Fe, respectively. The 937 and
1887~keV lines are both produced by $^3$He reactions on $^{16}$O.
\label{fig9}}
\end{figure}

As shown in Fig.~8, the $^3$He line at 1042 keV can also be
blended with another significant line, which is produced from the
reaction $^{56}$Fe($\alpha$,$pn\gamma_{1049}$)$^{58}$Co \cite{man99,koz02}.
Fig.~9b illustrates that the 1049~keV line should generally be 
considered, especially if the accelerated particle spectrum is hard (i.e. 
low values of $S$). In addition, the analysis of the 1042 keV line can be
complicated --as in this experiment-- by the proximity of the 1054 keV line
also produced in the reaction $^{16}$O($^3$He,$p$)$^{18}$F$^*$ (see Fig.~1 
and \S~III).

The 1081 and 1887~keV lines are expected to be separated from other
important lines (Fig.~8). They are, however, weaker than the 937 and 1042~keV
lines. The 1887~keV line is broader than the three other $^3$He lines,
because it is produced at higher $^3$He energies. Fig.~9c shows the
fluence ratio of the 1887 keV line to the 937 keV line, as a function of the
spectral index $S$. We see that the calculated ratio decreases by one order
of magnitude, from 0.4 for $S$=2 to 0.04 for $S$=6. Since both lines are 
produced from $^3$He-induced reactions on the same $^{16}$O target, this ratio 
is independent of the ambient medium and accelerated particle compositions.
It depends only on the accelerated $^3$He energy spectrum, which is 
important in understanding the acceleration mechanism in impulsive
solar flares (see \cite{pae03}). 

It is noteworthy that a significant synthesis of $^{18}$F nuclei in solar 
flares could also be identified through the characteristic, long living, 
511~keV emission expected from the $\beta^+$ decay of this isotope 
($T_{1/2}$=109.77~m). The detection of this delayed emission, maybe
with the {\it RHESSI} satellite, would furthermore provide a new insight 
into mixing and transport processes in the solar atmosphere. 

In summary, the four cross sections measured in this experiment could allow
one to determine with a good accuracy the abundance of $^3$He accelerated in
solar flares and interacting in the solar atmosphere. The strongest of the
gamma-ray lines from $^3$He-induced reactions is the one at 937 keV from the 
reaction $^{16}$O($^3$He,$p$)$^{18}$F$^*$, which however, can overlap with a 
line at 931 keV, produced in the reaction $^{56}$Fe($p$,$pn$)$^{55}$Fe$^*$. 
If observed, the 1887 keV line from the reaction
$^{16}$O($^3$He,$n$)$^{18}$Ne$^*$ would furthermore provide unique 
information on the energy spectrum of the accelerated $^3$He. 

\vspace{0.5cm}

\begin{acknowledgments}
We would like to thank the operator crew of the Orsay Tandem accelerator for 
their engagement in the preparation of the experiment, as well as 
Fr\'ed\'erico Garrido, Claire Boukari, Jacques Chaumont and Catherine Clerc 
for their assistance in the RBS measurements. 
\end{acknowledgments}


\end{document}